

\def\ifundefined#1{\expandafter\ifx\csname
#1\endcsname\relax}

\newcount\eqnumber \eqnumber=0
\def\beq{ \global\advance\eqnumber by 1 $$ }
\def\eeq{ \eqno(\the\eqnumber)$$ }
\def\label#1{\ifundefined{#1}
\expandafter\xdef\csname #1\endcsname{\the\eqnumber}
\else\message{label #1 already in use}\fi}
\def\(#1){(\csname #1\endcsname)}

\newcount\refno \refno=0
\def\[#1]{\ifundefined{#1}\advance\refno by 1
\expandafter\xdef\csname #1\endcsname{\the\refno}
\fi[\csname #1\endcsname]}
\def\refis[#1]{\item{\csname #1\endcsname.}}

\baselineskip=18pt
\magnification=1100
\def\tr{\;{\rm tr}\;}

\noindent {\bf ITP 93-41}

\noindent{\bf hep-th/9310002}
\vskip.5in

\centerline{{\bf THE UNTRUNCATED MARINARI-PARISI SUPERSTRING}}
\vskip 1in

\centerline{\bf G. Ferretti\footnote\dag{E-mail: ferretti@fy.chalmers.se }}
\vskip.5cm
\centerline{\sl Institute of Theoretical Physics}
\centerline{\sl  Chalmers University of Technology}
\centerline{\sl S-41296 G\"oteborg, Sweden}
\vskip 1in
\centerline{\bf Abstract}

It is shown that the bosonic angular degrees of freedom in the one dimensional
Marinari-Parisi superstring can be integrated out exactly in the Hamiltonian
formulation without having to perform the Dabholkar truncation.
The resulting Hamiltonian is that of a supersymmetric Calogero system
plus a four fermions interaction. This extra interaction vanishes for
all physical states with fermion number zero or one where supersymmetry is
manifest. We confirm that supersymmetry is nonperturbativly broken
by instanton effects.

\vfill\eject

\centerline{\bf a) The Marinari-Parisi superstring}

The Marinari-Parisi model \[marinari] describes a string moving in
one dimensional superspace. This is achieved by extending the well
know surface discretization techniques \[discretize], \[doublescale],
\[cequalone] to a supersurface.
There are many reasons why such generalization is desirable.
For one thing, one is interested in the non critical superstring
just as much as the bosonic one. Also, the supersymmetric formulation allows
one to study the $D=0$, pure gravity in a way that avoids the
instabilities present in the bosonic case. Finally, the model
provides a good example of supersymmetry breaking.

The action generating the dual triangulation of the world-sheet
embedded in one dimensional superspace is just the Wess-Zumino
model of a $N\times N$ supermatrix $\Phi$ with cubic interaction.
The components of the superfield are
\beq
    \Phi(t,\theta,\bar\theta)=X(t)+\Psi(t)\bar\theta+\bar\Psi(t)\theta +
    A(t)\bar\theta\theta.
\eeq
The bosonic fields $X$ and $A$ and the fermionic fields $\Psi$ and $\bar\Psi$
are $N\times N$ matrices and $\theta$, $\bar\theta$ are Grassmann numbers.
If we introduce the covariant derivative
\beq
    D=\bar\theta{\partial\over{\partial t}}+{\partial\over{\partial\theta}},
    \qquad
    \bar D=\theta{\partial\over{\partial t}}+
    {\partial\over{\partial\bar\theta}},
\eeq
and the potential
\beq
    W(z)={1\over 2}z^2 -{\lambda\over3}z^3,
\eeq
the desired action is
\beq
    S=\int dt d\bar\theta d\theta \tr\big(-{1\over 2}\Phi\bar D D\Phi
    +W(\Phi)\big).
\eeq

It is straightforward to integrate out the auxiliary field $A$ and to
perform the integration over the Grassmann variables $\theta$ and
$\bar\theta$ to obtain the expression for the supersymmetric action
in ordinary $D=1$ space-time:
\beq
    S={1\over 2}\int dt\tr\big(\dot X^2 + \bar\Psi\dot\Psi -
    \dot{\bar\Psi}\Psi + W^{'}(X)^2 - \bar\Psi W^{''}(X)\Psi
    - \bar\Psi\Psi W^{''}(X) \big),
\eeq
where, in writing down the last two terms under the trace, we used the fact
that the potential $W$ is only a cubic polynomial.

The Hamiltonian that follows from quantizing this one dimensional
system is the Hamiltonian of a supersymmetrical quantum mechanical system
\[witten]. It can be written explicitly in terms of the supercharges
$Q$ and $Q^\dagger$ as
\beq
    H={1\over 2}\{Q, Q^\dagger \},
\eeq
where
\beq
    Q=i\tr\Bigg(\bigg({\partial\over{\partial X}}-W^{'}(X)\bigg)a\Bigg).
    \qquad
    Q^\dagger=i\tr\Bigg(\bigg({\partial\over{\partial X}}+
    W^{'}(X)\bigg)a^\dagger\Bigg).
\eeq
We denote the fermionic creation and annihilation operator
valued matrices by $a^\dagger$
and $a$ and impose the following CAR on their components
$\{a^{\dagger i}_j, a^k_l\}=\delta^i_l \delta^k_j$.
Evaluating the anticommutator, one finds that the Hamiltonian consists
of a bosonic and a fermionic piece, both involving $N^2$ degrees of freedom:
\beq
    H=H_B+H_F,
\eeq
where
\beq
    H_B={1\over 2}\tr \bigg(-{{\partial^2}\over{\partial X^2}}+
    W^{'2}(X) - W^{''}(X) \bigg), \label{hbose}
\eeq
and
\beq
    H_F={1\over 2}
    \tr\bigg( a^\dagger W^{''}(X) a + a^\dagger a W^{''}(X) \bigg)=
    \tr\bigg(a^\dagger a - \lambda(a^\dagger X a + a^\dagger  a X)\bigg).
    \label{hfermi}
\eeq
Marinari and Parisi have analized the bosonic sector of the model \[marinari].
Neglecting the fermions allows one to integrate out the $N^2-N$ angular
degrees of freedom and to rewrite the Schr\"odinger equation in terms
on a gas of $N$ non interacting fermions \[brezin].
Away from the continuum limit, using a WKB approximation, one finds that
the vacuum energy is zero at weak coupling and scales like
$N^2(\lambda-\lambda_c)^{5/2}$ at strong coupling.

As $\lambda$ approaches its critical value the WKB energy vanishes
perturbatively. Formally, there is a wave function of zero energy given by
$\exp(-\tr W(X))|0>$. However, since the potential $W$ is cubic, this is
not a normalizable wave function, suggesting that there are nonperturbative
correction to the ground state energy. These corrections can be evaluated by
instanton calculus \[cargese] and found to be nonzero.
Supersymmetry is therefore broken by nonperturbative effects.

\centerline{\bf b) The Dabholkar truncation}

To estimate the non perturbative correction to the energy in a manifestly
supersymmetric way, one has to deal with states of
fermion number different from zero. (For example, one has to
compute the matrix element of the
supercharge $Q$ between the bosonic vacuum and its fermionic partner.)

The main problem when dealing with the full set of states is that the
supermatrix $\Phi$ can no longer be diagonalized by a unitary transformation.
Dabholkar \[dab] has shown that there is a consistent way of truncating the
theory that allows one to include some (but not all) states with non zero
fermion number while preserving manifest supersymmetry and reducing the
number of degrees of freedom from $2\times N^2$ to $2\times N$.

This is done by considering the unitary matrix $\Omega$ that diagonalizes the
bosonic matrix $X$ ($X = \Omega x \Omega^\dagger$)
and by defining the new fermionic
operator matrix $\hat a = \Omega^\dagger a \Omega$.
(A short remark: throughout the paper, we will make a slight abuse of
notation and denote both the diagonalized matrix and the ordered set
of eigenvalues by $x$; what is meant is always clear from the contest.)
One then considers only those fermionic states of the type
created by the diagonal elements of $\hat a$
\beq
    \hat a^{\dagger i_1}_{i_1} \hat a^{\dagger i_2}_{i_2}
    \hat a^{\dagger i_3}_{i_3} \cdots|0>,
\eeq
while excluding all non diagonal states.
On the subspace defined by this truncation, the supercharges can be written
in a way that is independent on the angular variables by introducing an
effective potential
\beq
    W_{\rm eff}(x_1\cdots x_N) = \sum_{i=1}^N W(x_i) +
    \sum_{i\not= j} \log |x_i-x_j|.
\eeq
The logarithmic factor in the effective potential comes from the expression
for the metric on the space of matrices when going from ``Cartesian'' to
``polar'' coordinates.

The explicit expression for the supercharge $Q$ in terms of the effective
potential is
\beq
    Q=i\sum_k\bigg({\partial\over{\partial x_k}} - {{\partial W_{\rm eff}}
    \over{\partial x_k}}\bigg)\hat a^{\dagger k}_k.
\eeq
The supersymmetric Hamiltonian $H=(1/2)\{Q, Q^\dagger\}$ is the
Hamiltonian of a supersymmetric Calogero system.
The consistency condition for this truncation to be physically meaningful is
that the subspace defined in this way is invariant under the action of the
supercharges $Q$ and $Q^\dagger$. This is indeed the the case for this model.
One can then carry on the double scaling limit of the charges and evaluate the
lifting of the vacuum energy. Dabholkar finds
\beq
    E_{\rm g.s.} \approx \kappa e^{- 2 S_{\rm inst.}/\kappa},
\eeq
$\kappa$ being the string coupling constant. Supersymmetry is therefore broken.
One can also give a collective field description of the theory \[collective].

\centerline{\bf c) The Untruncated theory}

As long as one is only interested in the ground state energy or in correlation
functions of operator that do not depend on the bosonic angular variables
(such as $\tr(aX^3)$ or $\tr(aX^4a^\dagger X^6)$), the truncated theory
gives all the informations that are needed. On the other hand, the
analysis of correlations functions for operators that are not in the singlet
sector is precluded from the very beginning. Such operators are of interest
since they create non trivial excitations and not much is known about
them\[gross].
Also, the above truncation, though consistent and supersymmetric, is
base dependent and it must be possible to write the theory is a way that does
not depend on the basis in the representation of $U(N)$.

Once again, one is facing the problem that it is impossible to diagonalize
simultaneously the bosonic and the fermionic degrees of freedom in the
Lagrangian. We will show however, that,
if one is willing to give up manifest supersymmetry, it is possible to
integrate out the angular degrees of freedom from the Hamiltonian!
The procedure used here is an adaptation of a  method already used in
the contest of Rajeev's ``Universal Yang--Mills Theory'' \[rajeev],
where one is also confronted with a matrix model with fermions \[ferretti].
One difference is that now, because of supersymmetry,
the fermions are in the adjoint representation
of $U(N)$ whereas in \[ferretti] they where in the fundamental one.

Let us begin our calculation by analyzing the symmetries of the most
general state vector of the Marinari-Parisi model. The most general
state of fermion number $n$ can be written as
\beq
    |\Psi(X)>=\Psi(X)^{i_1\cdots i_n}_{j_1\cdots j_n}
    a_{i_1}^{\dagger j_1}\cdots a_{i_n}^{\dagger j_n}|0>\equiv
    \Psi(X)\cdot (\otimes a^\dagger)|0>.
\eeq
where $\Psi(X)$ is a tensor antisymmetric under the exchange of the pairs
of indices $(i_k, j_k)$ and $(i_l, j_l)$ and $|0>$ is the fermionic vacuum.
(Do not confuse this $\Psi$ with the fermionic component of the
superfield.) From now on, we will use tensor notation.
The superfield $\Phi$ transforms under the adjoint representation of
$U(N)$, $\Phi\to U\Phi U^\dagger=Ad_U\Phi$ and therefore, all its
components $X$, $\Psi$, $\bar\Psi$, $A$ and the fermionic operators
$a$ and $a^\dagger$ will also transform under the adjoint representation.
Any state $|\Psi(X)>$ must be invariant under the action of $U(N)$, i.e.,
in terms of the tensor $\Psi(X)$,
\beq
    \Psi(U X U^\dagger) = (\otimes Ad_U\Psi)(X)
\eeq
Note that in this particularly simple case we can assume the above relation
to be imposed on the vectors themselves, not just on the rays.
Now recall that any hermitian matrix $X$ can be diagonalized by a
unitary matrix $\Omega$ as $X=\Omega x \Omega^\dagger$ and therefore we can
factor out the dependence of the tensor $\Psi$ from the angular variables as
\beq
    |\Psi(X)> =
    (\otimes Ad_\Omega\Psi)(x)\cdot (\otimes a^\dagger)|0>
\eeq
This is the most convenient form for further manipulations. From now on,
we think of the tensor $\Psi$
as depending, not on the matrix $X$, but only on the $N$
eigenvalues $x$ and express the dependence
on the angular variables through the action $\otimes Ad_\Omega$.

Before we carry on, let us recall that the
splitting $X=\Omega x \Omega^\dagger$
does not fix the representative of the state in a unique way. We can always
perform a subgroup of $U(N)$ transformations that leave $X$ invariant and
$x$ diagonal. This subgroup is clearly $U(1)^N\times S_N$. The subgroup
$U(1)^N$ consists of all the diagonal matrices in $U(N)$ and it acts on
$\Omega$ and $x$ in the following way:
\beq
    H\in U(1)^N: \qquad x\to x,\quad\hbox{and}\quad \Omega\to \Omega H.
\eeq
The subgroup $S_N$ is the discrete group of permutations of $N$ elements and
it acs by permuting the eigenvalues of $X$. $S_N$ can always be embedded in
$U(N)$ by considering $N\times N$ matrices $P$ with $N-1$ ``zeroes'' and $1$
``one'' in each row and column.
\beq
    P\in S_N: \qquad x\to P^\dagger xP,\quad\hbox{and}\quad \Omega\to\Omega P.
\eeq
This invariance must be imposed as a constraint on the tensor wave function.
Therefore, only those tensors that satisfy
\beq
    (\otimes Ad_H\Psi)(x)=\Psi(x)\quad\hbox{and}\quad
    (\otimes Ad_P\Psi)(x)=\Psi(PxP^\dagger)\label{constraints}
\eeq
are allowed as physical states. We will return to this point at a later stage,
for now let us proceed with the elimination of the angular variables.

Let us start by writing the expression of the metric on the space of
hermitian matrices in polar coordinates. In Cartesian coordinates, the
invariant lenght is just $ds^2=\tr(dX dX)$. Let us denote the $N$
radial variables as $x^i$ (as usual) and the remaining $N^2-N$ angular
variables by $\theta^a$. By setting $X=\Omega x\Omega^\dagger$ we obtain
\beq
    ds^2=\sum_{i=1}^N dx_i\otimes dx_i + \sum_{a,b=1}^{N^2-N}
    g_{ab}d\theta^a\otimes d\theta^b,
\eeq
where the angular part of the metric is given by
\beq
    g_{ab} = \sum_{k\not= l} (x_k - x_l)^2 \omega^{k*}_{la}(x,\theta)
    \omega^k_{lb}(x,\theta),
\eeq
and the complex functions $\omega^i_{ja}(x,\theta)$, $i,j=1\cdots N$,
$a=1\cdots N^2-N$ are defined implicitly through the relation
\beq
    (\Omega^\dagger d\Omega)^i_j =\omega^i_{ja}(x,\theta)d\theta^a
\eeq
and satisfy $\omega^{i*}_{ja} = -\omega^j_{ia}$.
The metric $g_{ab}$ has an inverse $g^{ab}$ satisfying the relation
$g^{ab}g_{ab} = N^2-N$. This fact, together with $S_N$ invariance implies
\beq
    g^{ab}\omega^{k*}_{la}(x,\theta) \omega^i_{jb}(x,\theta)=
    {1\over {(x_k-x_l)^2}}\delta^i_k \delta^l_j. \label{inverse}
\eeq
Also, it is well known that, setting $g=\det(g_{ab})$,
\beq
    \int d\theta \sqrt{g} = \prod_{i<j}(x_i-x_j)^2\equiv
    \Delta(x)^2. \label{vandermonde}
\eeq
Note that formula \(vandermonde) also follows, up to an
overall constant, from symmetry arguments and dimensional analysis.

Formulas \(inverse) and \(vandermonde) will allow us to integrate out
the bosonic angular variables from the energy.

The energy of a normalized state is given, in Cartesian coordinates, by
\beq
    E=\int dX <\Psi(X)|H_B + H_F|\Psi(X)> \equiv E_B + E_F,
\eeq
where the two Hamiltonians $H_B$ and $H_F$ are defined in \(hbose) and
\(hfermi).
The removal of the bosonic degrees of freedom from the fermionic energy
does not present any problem. One simply writes
\beq\eqalign{
    E_F&=\int dX <\Psi(X)|H_F|\Psi(X)>\cr =\int dx d\theta
    \sqrt{g}<0|(\otimes a)\cdot(\otimes Ad_\Omega \Psi(x))^\dagger
    &\tr\bigg( a^\dagger a -\lambda(a^\dagger Xa+a^\dagger aX ) \bigg)
    (\otimes Ad_\Omega\Psi)(x)\cdot(\otimes a^\dagger)|0>\cr
    =\int dx \Delta(x)^2 <0|(\otimes a)\cdot \Psi^\dagger(x)
    &\sum_{i,j=1}^N\bigg(1-\lambda(x_i + x_j)a^{\dagger i}_j a^j_i \bigg)
    \Psi(x)\cdot(\otimes a^\dagger)|0>.\cr}
\eeq
Now, just define a new tensor $\Xi(x)=\Delta(x)\Psi(x)$ satisfying the
invariance conditions that follow from the constraints \(constraints):
\beq
    (\otimes Ad_H\Xi)(x)=\Xi(x)\quad\hbox{and}\quad
    (\otimes Ad_P\Xi)(x)=(-1)^{\sigma(P)}\Xi(PxP^\dagger),
    \label{newconstraints}
\eeq
where $H$ and $P$ are the same as in \(constraints) and $\sigma(P)$ is the
parity of $P$.
The expression of the fermionic Hamiltonian on the states
$\Xi(x)\cdot (\otimes a^\dagger)|0>$, satisfying the constraints
\(newconstraints) is therefore given by
\beq
    H_F=\sum_{i,j=1}^N\bigg(1-\lambda(x_i + x_j)a^{\dagger i}_j a^j_i \bigg).
\eeq

The bosonic energy requires a more careful analysis due to the dependence of
the Laplacian on the angles $\theta^a$. We start by splitting the bosonic
energy $E_B$ in two part, a radial part, containing the radial part of the
Laplacian and the supersymmetric potential $W^{'2} - W^{''}$ and an angular
part, containing the angular part of the Laplacian.
\beq
    E_B = E_{\rm rad.} + E_{\rm ang.}.
\eeq
The radial part does not present any difficulty and it can be
rewritten at once as
\beq
    E_{\rm rad.}= \int dx \Delta(x)^2 <0|(\otimes a)\cdot \Psi^\dagger(x)
    {1\over 2}\sum_{i=1}^N \bigg(-{{\partial^2}\over{\partial x_i^2}}
    + W^{' 2}(x_i) - W^{''}(x_i)\bigg)\Psi(x)\cdot(\otimes a^\dagger)|0>.
\eeq
The angular part, on the other hand, is
\beq\eqalign{
    E_{\rm ang.}&={1\over 2} \int dx d\theta
    \sqrt{g} g^{ab}<\Psi(X)|{{\leftarrow\partial}\over
    {\partial\theta^a}}{{\partial\rightarrow}\over
    {\partial\theta^b}}|\Psi(X)>\cr=
    {1\over 2} \int dx d\theta
    \sqrt{g} &g^{ab}<0|(\otimes a)\cdot(\otimes Ad_\Omega\Psi)^\dagger(x)
    {{\leftarrow\partial}\over{\partial\theta^a}}
    {{\partial\rightarrow}\over{\partial\theta^b}}
    (\otimes Ad_\Omega\Psi)(x)\cdot(\otimes a^\dagger)|0>.\cr}
    \label{temporary}
\eeq
Using the fact that
\beq
    {\partial\over{\partial\theta^a}}Ad_\Omega=
    Ad_\Omega\;ad_{\Omega^\dagger{\partial\over{\partial\theta^a}}\Omega},
\eeq
and that the generators of the adjoint action on the state
can be expressed in terms of fermionic bilinears
\beq
    ad_{\Omega^\dagger{\partial\over{\partial\theta^a}}\Omega}=
    \omega_{ak}^l(a^{\dagger p}_l a^k_p - a^{\dagger k}_p a^p_l),
\eeq
we rewrite \(temporary) as
\beq\eqalign{
    E_{\rm ang.}&={1\over 2} \int dx d\theta
    \sqrt{g} g^{ab} \omega ^{l*}_{am}\omega^i_{bj}\cr &
    <0|(\otimes a)\cdot\Psi^\dagger(x)
    (a^{\dagger l}_p a^p_m - a^{\dagger p}_m a^l_p)
    (a^{\dagger q}_i a^j_q - a^{\dagger j}_q a^q_i)
    \Psi(x)\cdot(\otimes a^\dagger)|0>.\cr}
\eeq
Using \(inverse) and \(vandermonde), we can now integrate out the
angular dependence altogether:
\beq\eqalign{
    E_{\rm ang.}&={1\over 2} \int dx \Delta(x)^2
    \sum_{l\not= m}{1\over{(x_l-x_m)^2}}\cr &
    <0|(\otimes a)\cdot\Psi^\dagger(x)
    (a^{\dagger l}_p a^p_m - a^{\dagger p}_m a^l_p)
    (a^{\dagger q}_l a^m_q - a^{\dagger m}_q a^q_l)
    \Psi(x)\cdot(\otimes a^\dagger)|0>.\cr}
\eeq
The effect of the integration over the angular variables is the introduction
of a (non manifestly supersymmetrical) four fermi interaction. We stress that
this is an exact equivalence and it is valid on all the states without
any truncation. It is not surprising that this expression is not manifestly
supersymmetric because we have integrated out some of the bosonic degrees
of freedom.

We can now carry out the same substitution $\Psi(x)\to \Xi(x)$ and it is well
known \[brezin] that no extra terms are generated. The bosonic Hamiltonian
acting on the states $\Xi(x)\cdot (\otimes a^\dagger)|0>$ is therefore
obtained by adding the radial and the angular contributions;
\beq
    H_B={1\over 2}\sum_{i=1}^N \bigg(-{{\partial^2}\over{\partial x_i^2}}
    + W^{' 2}(x_i) - W^{''}(x_i)\bigg)+
    {1\over 2}\sum_{l\not= m}{1\over{(x_l-x_m)^2}}
    (a^{\dagger l}_p a^p_m - a^{\dagger p}_m a^l_p)
    (a^{\dagger q}_l a^m_q - a^{\dagger m}_q a^q_l).
\eeq

We conclude this section by rewriting the physical constraints
\(newconstraints) in a more explicit form and specializing, as an illustration,
to the fermion numbers zero and one. For an arbitrary element $T$ of the
Lie algebra of $SU(N)$, we have seen how to write the generators
of $\otimes ad_T$ in terms of fermionic bilinears. Specializing to the
Cartan subalgebra, we can rewrite the first constraint in \(newconstraints) as
\beq
    \sum_{p,\hbox{ not } i}(a^{\dagger p}_i a^i_p -
    a^{\dagger i}_p a^p_i)\Xi(x)\cdot
    (\otimes a^\dagger)|0>,\quad\hbox{for all $i=1\cdots N$}.
\eeq
The second constraint implies the impossibility of factorizing the state
vector as $f(x) |\hbox{fermi}>$ (where $f$
is a scalar) except when $|\hbox{fermi}> = |0>$.
In particular, the zero fermions wave function is given as $\chi(x)|0>$
in terms of a totally antisymmetric scalar function $\chi(x)$,
whereas the one fermion wave function is given as
\beq
    \sum_m \hat\chi(x_m,x_2,\cdots,x_{m-1},x_1,x_{m+1}\cdots,x_N)
    a^{\dagger m}_m |0>, \label{onefe}
\eeq
$\hat\chi$ being antisymmetric in the last $N-1$ variables.
Note that all the zero and one fermion states belong to the
Dabholkar truncated sector. This is obvious if one realizes that all
the states in the truncated sector trivially satisfy the first of
\(newconstraints). However, not all the states in the truncated
sector satisfy the second of \(newconstraints); only those of the form
\(onefe) are allowed as true physical states. To find physical states that do
not belong to the truncated sector, one has to go to fermion occupancy
equal to two. There, one can construct  states like
$\Xi(x)^{ij}_{kl} a^{\dagger l}_i a^{\dagger k}_j |0>$ that satisfy
\(newconstraints) but do not belong to the truncated sector.

As far as the calculation of the vacuum energy goes, the same results
described in \[dab] apply here, mainly because
the instanton calculation only involves states
with fermion number zero and one. Hence, we see that supersymmetry is
broken nonperturbativly.

The obvious (but important)
thing to notice though is that the part of the Hamiltonian that is
quartic in the fermionic operator vanishes on all states of fermion
number zero or one. This means that supersymmetry is manifest on those states
where the Hamiltonian reduces to that of a supersymmetric Calogero system
with the {\it same} potential $W$ as in the original action and not the
effective potential described in \[dab] which is ultimately
arising from the truncation. The calculation of the nonperturbative corrections
to the vacuum energy proceedes just as in \[dab], without even having to worry
about the effective potential. One finds once again that
$E_{\rm g.s.} \approx \kappa \exp(- 2 S_{\rm inst.}/\kappa)$.

It would be interesting to investigate the effect of the quartic Hamiltonian
on the states with higher fermion number. The explicit form
of the Hamiltonian and of the states is simple enough to allow some
investigations on the non singlet sector. This should be easier to do
here than in the case of the bosonic string on the circle, let alone
strings in higher dimensions.

The model described in this paper also suggests a possible generalization
of the Calogero systems \[calogero] in a direction that has
not yet been explored.
It might still be possible to solve the theory exactly even with this
quartic term.

\centerline{\bf Acknowlegements}

Most of the group theory techniques used here were explained to me by
S.G. Rajeev while we were working on a similar problem. This work was
initiated while visiting S.I.S.S.A., whose hospitality is greatfully
acknowledged. Finally, I want to thank L. Brink and B. Nilsson for
discussions.

\vfill\eject

\centerline{\bf References}

\refis[marinari] E. Marinari and G. Parisi, Phys. Lett. {\bf B 240} (1990) 375.
\refis[discretize]
           F. David, Nucl. Phys. {\bf B 257} (1985) 45;
            \hfill\break
           F. David, Nucl. Phys. {\bf B 257} (1985) 543;
            \hfill\break
           V.A. Kazakov, Phys. Lett. {\bf B 150} (1985) 28;
            \hfill\break
           J. Ambj\o rn, B. Durhuus and J. Fr\"o hlich, Nucl. Phys.
             {\bf B 257} (1985) 433.
\refis[doublescale]
           E. Brezin and V. Kazakov, Phys. Lett.
           {\bf B 236} (1990) 144;
           \hfill\break
           M. Douglas and S. Shenker, Nucl. Phys. {\bf B335} (1990) 635;
           \hfill\break
           D. Gross and A. Midgal, Phys. Rev. Lett. {\bf 64} (1990) 127;
\refis[cequalone]
           D.J. Gross and N. Miljkovic, Phys. Lett. {\bf B 238} (1990) 217;
           \hfill\break
           E. Brezin, V.A. Kazakov and A.B. Zamolodchikov, Nucl. Phys.
           {\bf B 338} (1990) 673;
           \hfill\break
           P. Ginsparg and J. Zinn-Justin, Phys. Lett. {\bf B 240} (1990) 333.
\refis[witten] E. Witten, Nucl. Phys. {\bf B 185} (1981) 513
               \hfill\break
               E. Witten, Nucl. Phys. {\bf B 202} (1983) 253.
\refis[brezin] E. Brezin, C. Itzykson, G. Parisi and J.B. Zuber,
               Comm. Math. Phys. {\bf 59} (1978) 35.
\refis[cargese] G. Parisi, in Workshop on Random surfaces, quantum
                \hfill\break
                gravity and strings, Carg\`ese, France 1990;
                \hfill\break
                S. Shenker, in Workshop on Random surfaces, quantum
                \hfill\break
                gravity and strings, Carg\`ese, France 1990.
\refis[dab] A. Dabholkar, Nucl. Phys. {\bf B 368} (1992) 293.
\refis[collective]
                   A. Jevicki and B. Sakita, Nucl. Phys. {\bf B 165}
                   (1980) 511;
                   \hfill\break
                   S.R. Das and A. Jevicki, Mod. Phys. Lett., {\bf A 5}
                   (1990) 1639;
                   \hfill\break
                   A. Jevicki and J.P. Rodrigues, Phys. Lett. {\bf B268}
                   (1991) 53.
                   \hfill\break
                   J.D. Cohn and H. Dykstra, Mod. Phys. Lett. {\bf A 7}
                   (1992) 1163;
                   \hfill\break
                   J.P. Rodrigues and A.J. Van Tonder, Int. J. Mod. Phys.
                   {\bf A 8} (1993) 2517.
\refis[gross] D. Gross and I.R. Klebanov, Nucl. Phys. {\bf B344} (1990) 475;
              \hfill\break
              P. Marchesini and E. Onofri, J. Math. Phys. {\bf 21} (1980) 1103.
\refis[rajeev] S.G. Rajeev, Phys. Rev. {\bf D 42} (1990) 2779;
               \hfill\break
               S.G. Rajeev, Phys. Rev. {\bf D 44} (1991) 1836.
\refis[ferretti] G. Ferretti and S.G. Rajeev, Phys. Lett.
                 {\bf B 244} (1990) 265;
                 \hfill\break
                 G. Ferretti, Phys. Lett. {\bf B 284} (1992) 325.
\refis[calogero] F. Calogero, J. Math. Phys., {\bf 10} (1969) 2191;
                 \hfill\break
                 F. Calogero, J. Math. Phys., {\bf 10} (1969) 2197;
                 \hfill\break
                 F. Calogero, J. Math. Phys., {\bf 12} (1971) 419;
                 \hfill\break
                 L. Brink, T.H. Hansson and M.A. Vasiliev,
                 Phys. Lett., {\bf B 286} (1992) 109.
\bye